\documentclass[USenglish]{llncs}

\usepackage[utf8]{inputenc}

\usepackage{siunitx}
\usepackage{graphicx}
\usepackage{thmtools,thm-restate}

\usepackage[adversary,probability,sets,operators,landau]{cryptocode}
\renewcommand{\sample}{\xleftarrow{\$}}

\newcommand{\expsa}{\textsf{\textup{EXP-SA}}}

\usepackage{amsmath}
\usepackage{amssymb}

\begin{document}
\title{Safety in Numbers: Anonymization Makes Centralized Systems Trustworthy}

\author{Lachlan J.~Gunn \and Andrew Allison \and Derek Abbott}
\institute{School of Electrical and Electronic Engineering \\
		The University of Adelaide, Australia \\ %
\email{\{lachlan.gunn, andrew.allison, derek.abbott\}@adelaide.edu.au} }

\maketitle

\begin{abstract}{%
Decentralized systems can be more resistant to
operator mischief than centralized ones, but they are substantially harder to
develop, deploy, and maintain.  This cost is dramatically reduced if the
decentralized part of the system can be made highly generic, and thus
incorporated into many different applications.  We show how existing anonymization
systems can serve this purpose, securing a public database against
equivocation by its operator without the need for cooperation by the
database owner.  We derive bounds on the probability of successful
equivocation, and in doing so, we demonstrate
that anonymization systems are not only important
for user privacy, but that by providing privacy to machines
they have a wider value within the internet infrastructure.
}\end{abstract}

\section{Introduction}
Suppose you want to call someone, but do not
know their phone number.  How do you find it?  The obvious way is to look them up
in a phone book, but the phone company might have placed a different
address under their name.  If they are particularly security conscious, then
you might presume that, when they received their phone book, they checked
to see whether their number is correctly listed.  But what if it were not modified
in every phone book, but some contain the real number and some contain
a false one?

The authentication of database entries and user attributes is an important problem
in information security; one of the most prominent applications is in key-distribution for
end-to-end secure messaging.  Some systems use centralized key-distribution services,
placing trust in the operators of their servers.  Others use decentralized methods,
but existing methods come with their own limitations; the public-key infrastructure
allows most certificate authorities to impersonate anyone, and mainstream blockchain
systems waste power calculating proofs of work.
The result is that even when a database can be realistically distributed, the designers
of many systems choose not to do so.

This task is greatly simplified if we can decentralize a system in a generic
way, adding standard components that can be reused for many systems.  We show
in Sections~\ref{sec:protocol} and~\ref{sec:analysis} that an anonymization system
serves this purpose, in many cases without modification or even the cooperation
of the central server.

Our contribution in this paper is to show that a variation of the
multi-path probing~\cite{wendlandt-tofu} approach
used by \emph{DetecTor}~\cite{detector} is provably secure.
Users simultaneously make identical
requests to a central service via an anonymizer.  If they receive consistent
responses, then they can assure themselves that the server provides
identical responses to identical requests; we show in
Section~\ref{sec:broadcast} that a server can successfully equivocate across
$N$ users for $M$ rounds with probability at most $N^{1-M}$.

This approach has a number of advantages over other anti-equivocation techniques in the literature:
\begin{itemize}
	\item \textbf{No bootstrapping problem.}  By using an existing anonymity system to
		audit quite general services, new systems can obtain the benefits
		of distributed auditing without an existing community to provide operator-diverse
		monitoring systems.
	\item \textbf{Scalability.}  Users do not need to communicate with each other,
		except to signal that the service has misbehaved.  As a result, the communication
		overhead is only $\bigO{\log \epsilon}$ for a given security level $\epsilon$.
	\item \textbf{Computational efficiency.}  Because we do not use a proof-of-work
		system, no computational power is wasted on what is generally pointless busywork
		whose only purpose is to make participation costly.
		
		This is relevant to our first point: a new proof-of-work system is not
		secure until it has enough miners to out-compute any potential attacker.  This
		creates a chicken-and-egg problem, in that the system is not secure until it is
		widely adopted, which will not happen if it is insecure.
	\item \textbf{No server-side cooperation needed.}  This approach does not
		require any changes on the server-side; as a result, it is quite practical for
		motivated users to audit existing services without the need for effort or
		cooperation from their operators.
\end{itemize}
The first of these points is particularly important, as many pieces of
software begin as a small-scale project by individuals
or groups without third-party commitment.  Our protocol provides a distributed
auditing capability that has until now been completely unavailable to such projects.

\subsection{Motivation}
Our principal motivation for the development of this auditing method is to allow
the use of centralized key-distribution servers in a secure manner.
Key distribution is a difficult problem to solve, and as it stands there are few solutions
that do not centralize trusted operations to a significant degree, requiring manual
verification on the part of users in order to eliminate the risks posed by malicious
infrastructure operators.

The need for manual effort is problematic in multiple ways; the first is that most users
will simply not bother, but even amongst those users who do make the effort, they
will not necessarily wait for the verification to take place before communicating.  This
leaves a window of vulnerability before an attack is detected, which in the case of
manual in-person verification may be very long indeed.

Our desire, then, is to allow users to take responsibility for the security of their
own identity to the greatest extent possible, but in a way that does not require
a significant degree of manual effort.

\subsection{Public-key distribution: the \emph{status quo}}
The most widely-accepted systems that are configured by end-users,
such as Secure Shell (SSH)~\cite{rfc4251} and WhatsApp~\cite{whatsapp-crypto},
tend to use a trust-on-first-use~\cite{wendlandt-tofu}
model, in which initial communication takes place with either no
or only manual authentication, after which the the user is alerted 
to key changes.  However, this does not prove the identity of the user
unless the two parties use some out-of-band authentication method.

This can be overcome by standards such as X.509~\cite{rfc5280},
which use signed certificates to verify identities:
X.509 is widely adopted by email clients, but
the need to acquire certificates from a commercial certificate authority
has prevented it from seeing any significant use; in addition, most certificate
authorities can issue certificates for anyone in the world, an ability that has
been abused on a number of occasions both by attackers and the operators of
the authorities.
Pretty Good Privacy (PGP)~\cite{rfc4880} aims to provide message-level security to the
masses, but has been hampered by the difficulty of its key management,
which depends upon personal contact to establish trust relationships.

Identity-based cryptography provides another approach, in which
trust in a key needs only be established at the organizational level rather than
between individual users, but this allows access to private keys by service providers;
the ubiquity of adversaries with coercive powers and an interest in mass surveillance
means that this is entirely inadequate from a privacy standpoint.  While the risk might be
mitigated with the aid of threshold cryptosystems or other distributed
approaches, if the desire to decrypt a user's communications exists at an
organizational level then threshold decryption and secret sharing provides
little protection to users.

\subsection{Related work}

The problem of obtaining agreement on a value amongst several---possibly
malicious---users is
an old one, known as the Byzantine Generals problem, and was first analyzed
by Lamport \emph{et al.} in 1982~\cite{byzantine-generals}.  Several officers
plan for an attack, in which they must act simultaneously
in order to be successful.  This is complicated by the knowledge that some of
the officers may be traitors---including the general in command of all of
them---and may therefore send different messages to different units in an
effort to induce a doomed attack.

Consensus protocols have seen increasing prominence in recent years
with the rise of cryptocurrencies such
as Bitcoin~\cite{bitcoin}, whose security against double-spending depends
upon public scrutiny of the submitted transactions.  If consensus on the
transaction ledger is broken---that is to say, if different users see different
values---different transaction records can be sent to different
users, allowing double-spending to occur.

\subsubsection{Traditional consensus protocols}

Traditional consensus-based approaches~\cite{lynch-distributed-algorithms}
are effective, but typically do not scale well to large numbers of
participants~\cite{vukolic16}.  Significantly, they require communication channels
between many of the nodes taking part in the consensus protocol, with
resistance to traitors being limited by the connectivity of the network graph.
This is inconvenient in practice, as it requires individual clients to discover
and communicate with large numbers of independent nodes, and requiring
a large community in the first place in order to bootstrap the network, since
additional nodes controlled by the same operator make the system
\emph{less} secure as it increases the number of traitorous nodes if the
the operator is malicious.

\subsubsection{Proof-of-work protocols}

The Bitcoin protocol~\cite{bitcoin} prevents the consensus from being split
by requiring a proof-of-work in order for the transaction to be published,
hash-linked to the previous state of the ledger.  This functions somewhat
analogously to a voting system, with the state of the ledger being collectively
determined by whichever group has the most computational power.

This type of protocol has the advantage over classical protocols~\cite{cachin00}
of not requiring large amounts of communication amongst the users in question.
For example, the algorithm presented in~\cite{cachin00} requires
a message count that is $\bigO{N^2}$ in the number of users.


A disadvantage of the proof-of-work approach is the need for
popularity---the security of proof-of-work-based systems comes from the expense
of performing enough computation to compete with the rest of the network, meaning
that smaller projects will initially be completely controlled by their founders, and even
after the appearance of independent miners, they will be vulnerable for some time to the sudden
appearance of an adversary with large amounts of computing power.  Furthermore,
the computation of these proofs-of-work requires large amounts of power, making the
scheme rather inefficient.

\subsubsection{Collective signing}
An alternative method has recently been proposed
that uses collective signing~\cite{syta-cosi,kogias-bitcoin-collective-signing}.  This
allows consensus to be efficiently demonstrated by collectively-generated digital
signatures.  If the consensus group is known in advance, then this allows us to
ensure that the entire group has accepted the same piece of data.

Knowledge of the group members is a potential problem;
in~\cite{kogias-bitcoin-collective-signing}, where the collective signing approach was
applied to Bitcoin, group membership is given to those who have recently mined a block,
taking advantage of the proof-of-work system to prevent Sybil attacks.

Without a proof-of-work system, some other transparent way of determining who will be invited to
take part in the collective signature process is necessary.  Nonetheless, such an
approach will prove effective, if the necessary infrastructure comes into being.

The primary disadvantage of this approach is the need for dedicated verification
infrastructure; this creates a bootstrapping problem when new types of verification
are needed.  Nonetheless, this may be overcome for verification tasks such as
domain verification that have wide commercial appeal.

\subsubsection{Multi-path probing}

Our proposed technique is a special case of multi-path probing.  Multi-path probing
involves accessing a service from several points of view in order to detect local variations
in responses such as caused a man-in-the-middle attack located far from
the service in question.  

The first such system was \emph{Perspectives}~\cite{wendlandt-tofu}.  This system
uses a number of \emph{notary servers}, which scan publicly-accessible web services for
keys.  By doing so regularly, they obtain a record of a service's public-key history, and thus
allowing users to convince themselves that the server has not changed its public key
recently.  By accessing multiple such notaries, they can see the key as it appears from
several network perspectives.  This reduces the risk posed by a malicious notary.
Unfortunately, much of this functionality depends upon knowledge of the protocol in
use, so that the public key can be extracted.  This means that new services cannot
be audited with Perspectives until they have developed a following sufficient to justify
their support by a large number of notary servers.

A simpler approach, and the direct inspiration for our scheme, is implemented by
\emph{DoubleCheck}~\cite{doublecheck}.  When connecting to a server, DoubleCheck
makes a second connection via Tor, which it uses to acquire a copy of its certificate.  This
certificate is compared with that obtained via the direct connection, and the user is warned
if they do not match.  The same approach is used by \emph{DetecTor}~\cite{detector},
which is notable for suggesting that operators use it to verify the state of \emph{their own}
servers.

The CONIKS~\cite{coniks} directory system, includes a Perspectives-like scheme in its architecture,
going so far as to include bounds on the probability of successful equivocation by a
given number of malicious auditors.  Their analysis is related to a special case of our own,
but crucially assumes the existence of independent auditors who store and distribute their signed
tree roots.  This allows clients to see the database from multiple viewpoints, but creates a
bootstrapping problem.

Our contribution is to demonstrate that it is possible to design an anti-equivocation system
like that proposed for CONIKS without dedicated auditing systems.  We describe a
DetecTor-like system whose consensus is provably secure, relative to the sender
anonymity of the anonymization system in use.

\section{Verification protocol}\label{sec:protocol}

\begin{figure}
	\centering
	\includegraphics{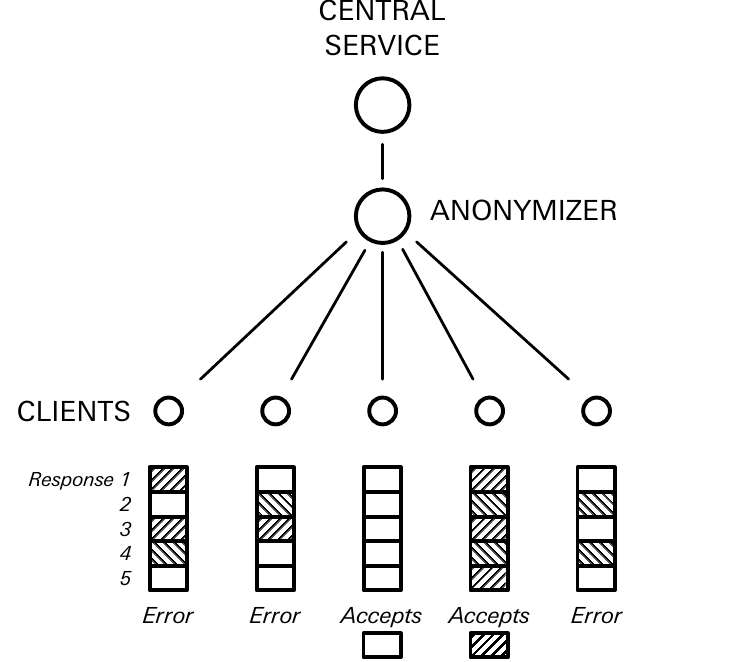}
	\caption{Interpretation of the results obtained from the protocol.  Clients that
			have not received consistent responses from the server reject
			the response from the server, which they know to be faulty.
			Clients that have consistently received the same response accept
			it as unequivocated.  In this figure, the server has equivocated,
			with the third and fourth clients being unaware of the fact and the
			others detecting the misbehavior of the service.}
	\label{fig:results-diagram}
\end{figure}

Suppose that Bob wishes to acquire a piece of information from an untrusted anonymously-accessed
service, and Alice the auditor can detect whether a given response from the server is valid.
The protocol that we propose is as follows:
\begin{enumerate}
	\item At a predetermined time, Alice and Bob both request a copy of the message
		from the service.
	\item The service responds to their requests with the message provided.
	\item Steps one and two are repeated $M$ times.
	\item If Bob does not receive $M$ identical responses, he publicly signals an error.
	\item Alice checks that the messages that she has received are identical and valid,
			and publicly signals an error if not.
\end{enumerate}
We show this in Figure~\ref{fig:results-diagram}.  Clients who see evidence of equivocation
know that the service is untrustworthy, and can report its misbehavior.  If the responses
are signed, these clients can prove to third parties that the server has equivocated,
providing a substantial deterrent to misbehavior on the part of the service.

Any anonymizing system can be used for this protocol, but in general synchronized
system such as a mix-net will be more effective, as these provide little-to-no room for
timing attacks.  In practice, low-latency anonymization networks such as Tor are far
more available than mix-nets; we discuss the methods used to close the timing
side-channel attacks in Appendix~\ref{sec:impl}.

\section{Security Analysis}\label{sec:analysis}

There are many anonymizing systems in use,
the most popular by far being Tor~\cite{tor-design}.
One of the goals of Tor is
to prevent users from being deanonymized over the long
term~\cite{tor-design}.  This is a reasonable target, given that one of Tor's
stated purposes is the protection of dissidents and journalistic sources
from state-level adversaries.  Compromizing a single request over the course
of many years might well result in catastrophic consequences for the user; even
if that single request does not contain any compromising information, it may tie
them to a pseudonym---e.g. a social media account whose activities are known.
As an example of this, the head of the hacking group \emph{LulzSec} was
arrested after connecting to an online chat server on just a single occasion
without using Tor~\cite{register-lulzsec}.

Our requirements are different---whereas a dissident, leaker, or criminal
desires to minimize the probability that they will ever be deanonymized,
our desire is to minimize the probability that an individual request is
deanonymized, since the security of the
design that we will describe shortly is determined by the number
of requests that can be made without being connected to one another.
We will discuss this distinction in greater detail in Section~\ref{sec:adversary},
but it is important to highlight that what we describe is only one of many
possible definitions of anonymity that has been chosen to meet our needs.

\subsection{Definitions}\label{sec:notation}

We begin by defining some notation.  We consider a set of users
$\mathcal{U} = \{U_1, \ldots, U_N\}$ who take part in the protocol above.
This is our anonymity set.

We write the set of injections from $A$ into $B$ as $\mathrm{Inj}(A \rightarrow B)$,
and bijections from $A$ to $B$ as $\mathrm{Bij}(A \rightarrow B)$.

These users connect to a service via an anonymizer, all making identical requests.  We model
this process in Figure~\ref{fig:anonymizer}.  The anonymizer makes a request to the adversary
on behalf of the clients, providing partial information on which response will go to which user.

\begin{figure}
\centering
\fbox{\procedure{$\textsc{AnonRequest}(\mathcal{U}, \adv, \mathcal{L})$}{
	\pccomment{Select request identifiers by random assignment.} \\
	R_\mathrm{I}(\cdot)  \sample \mathrm{Bij}(\mathcal{U} \rightarrow \ZZ_{|\mathcal{U}|}) \\ \\
	\pccomment{The adversary provides a response for} \\
	\pccomment{each request number} \\
	R_\mathrm{V}(\cdot) \sample \mathcal{A}(\mathcal{L}(R_\mathrm{I})) \t \pccomment{$R_\mathrm{V} : \ZZ_{|\mathcal{U}|} \rightarrow \bin^*$} \\  \\
	\pccomment{Return the response identifiers and values.} \\
	\pcreturn R_\mathrm{I}(\cdot), R_\mathrm{V}(\cdot)
}}
\caption{A model of an anonymously-accessed service, where $\adv$ is the
			potentially-malicious service, and $\mathcal{L}$ is a leakage function
			that captures the information leaked to the adversary.  In the case of
			Tor, for example, $\mathcal{L}$ is the user-to-request mapping
			$R_\mathrm{I}$ with its domain restricted
			to users whose entry guards are surveilled by the attacker.
			The service accepts a set of users, and selects a random mapping
			from users to request identifiers.  The adversary is given system-dependent
			partial information on the source of each request, and invited to provide a
			response to each request.}
\label{fig:anonymizer}
\end{figure}

\subsection{Adversary model}\label{sec:adversary}

We define our security relative to the security of an anonymity system,
and in particular to the notion of sender anonymity as defined by Pfitzmann and
K\"{o}ntopp~\cite{pfitzmann-anonymity}, and
loosely follow the formalisation given by Backes~et~al.~\cite{backes13},
but extended to $N$ simultaneous
users and $M$ request-response rounds. 
We select this definition because it provides the
most direct route to our statistical quantities of interest; this type of definition
can be related to indistinguishability-based definitions such as those by
Backes et~al.\@ in a straightforward manner~\cite{backes13}.

\begin{definition}[Sender-anonymous service]\label{def:anonymity}

Suppose a set of users $\mathcal{U} = \{U_1, \ldots, U_N\}$ each make a series of $M$
identical and synchronous requests to a service via an anonymizer,
receiving a response, as in Figure~\ref{fig:anonymizer}.

Then, consider the experiment in
Figure~\ref{fig:sa-experiment} for any adversary $\adv$, with the leakage function
$\mathcal{L}$ being
a system parameter.  We call the combination of anonymization system and service
\emph{$\epsilon$-sender-anonymous}, $\epsilon \ge 0$, if for all adversaries $\adv$,
\begin{align}
	\prob{\expsa_{\mathcal{L},\mathcal{U},M}(\mathcal{A}) = 1} \le \frac{1}{N!}\left(1+\epsilon\right) .
\end{align}
\begin{figure}
\centering
\fbox{
\procedure{$\expsa_{\mathcal{L},\mathcal{U},M}(\mathcal{A})$.}{
	\pccomment{Prime the adversary with $M-1$ anonymous requests.} \\
	\pcfor i = 1 \ldots (M-1) \\
	\t 	\mathrm{State} \sample State \parallel \textsc{AnonRequest}(\mathcal{U}, \adv_\mathrm{State}, \mathcal{L}) \\
	\pcendfor \\
	\pccomment{Perform the final request.} \\
	R_\mathrm{I}(\cdot), R_\mathrm{V}(\cdot) \sample \textsc{AnonRequest}(\mathcal{U}, \adv_\mathrm{State}, \mathcal{L}) \\
	\pccomment{Let the adversary identify a response identifier for each user.} \\
	\hat{R}(\cdot) \sample \adv(\mathcal{U}, \mathcal{L}, \mathrm{State}) \\
	\pccomment{The adversary wins if they sent their responses to the} \\
	\pccomment{users that they thought.} \\
	\pcif \hat{R}(\cdot) = R_\mathrm{I}(\cdot) \\
		\t\pcreturn 1 \\
	\pcelse \\
		\t \pcreturn 0 \\
	\pcendif
}}
\caption{Security experiment for sender-anonymity.  An anonymity system,
	defined by its leakage function $\mathcal{L}$, is used to make requests to
	an adversary who aims provide particular messages to particular
	users.  The adversary is asked to determine the users to whom each of
	its responses were sent; it wins if it correctly identifies all of the recipients.}
\label{fig:sa-experiment}
\end{figure}
\end{definition}

This definition assumes that all users operate in lockstep,
masking their identities by making identical requests with covert channels
sufficiently masked that the probability of successfully linking
consecutive requests is no better than chance.  We use a multiplicative
parameter $1+\epsilon$ rather than an additive one because this simplifies
the analysis to follow; the same results hold with an additive parameter
$\epsilon_+ = N! \epsilon$.

The most straightforward
way to achieve this is the mix-net~\cite{chaum-mix}, where a chain of
hosts, called \emph{mixes}, re-encrypt and shuffle fixed-sized messages,
guaranteeing anonymity so long as at least one member of the chain is honest.  The
anonymity set here is the set $\{U_i\}$ of users who take part in the protocol.

From our perspective, this means that the adversary
is unable respond so in such a way as to target a particular user with a particular
response.  Whether the adversary has
compromised the service or is performing a man-in-the-middle attack
is immaterial; all we require is that they cannot
deanonymize the requests in time to send messages tailored to a particular
user.

In some systems this is proven with respect to particular computational hardness
assumptions~\cite{young-drunk-motorcyclist}, whereas other systems such as Tor
are \emph{ad-hoc}~\cite{camenisch-onion} and will fall to a global passive adversary.
Our approach is implicitly conditional upon whichever assumptions are
made by the underlying anonymization system; should a provably-secure alternative
to Tor become equally widespread, it will serve just as well.

There exists the possibility that an attacker might use a denial-of-service to prevent
an individual user from accessing the server, if the attacker is able to identify the link
that they use to connect to the anonymizing service.  Defeating this type of attack
is outside the scope of this paper, however we note that it will always be recognized
as a fault by the user in question and reported as such.

While this definition makes clear the capabilities of the adversary, it is
not ideal for calculation.  We will thus make extensive use of the following
theorem:

\begin{theorem}\label{thm:response-distribution}
Consider the protocol from Section~\ref{sec:protocol} with $N$ users, where the anonymizing service
is a synchronous $\epsilon$-sender-anonymous service, as in Definition~\ref{def:anonymity}.
Then, for any adversary $\adv$ with arbitrary knowledge of the recipients of the previous
messages, the recipients of responses $1, \ldots, N$ are approximately
uniformly distributed over $\mathrm{Bij}(\ZZ_N \rightarrow \mathcal{U})$,
with each of the $N!$ mappings from responses to recipients occuring with probability at most
$(N!)^{-1}(1 + \epsilon)$.

\begin{proof}
Consider an arbitrary round of requests in the described protocol.
We note that the security experiment in Figure~\ref{fig:sa-experiment} mirrors steps
one to three of the protocol, with the function $R_\mathrm{I}(\cdot)$
representing the response destinations for the round under consideration.  Thus, by
our assumption of $\epsilon$-sender-anonymity, the adversary can predict all of the
response destinations with probability at most $(N!)^{-1}(1 + \epsilon)$. 

We now proceed by contradiction.  Suppose the adversary can act in such a way that
some response-user mapping $R: \ZZ_N \rightarrow \mathcal{U}$ occurs with
probability greater than $(N!)^{-1}(1 + \epsilon)$.  Then, in
Figure~\ref{fig:sa-experiment}, the adversary can select this mapping as their
prediction $\hat{R} : \ZZ_N \rightarrow \mathcal{U}$ of the message destinations.
By supposition, this is is correct with a probability greater than $(N!)^{-1}(1 + \epsilon)$,
in contradiction of Definition~\ref{def:anonymity}, yielding the desired contradiction.
\end{proof}
\end{theorem}

We note that when $\epsilon = 0$, this implies that the recipients of each
message are perfectly uniformly distributed.

It is this mixing property that we use to provide security.  Any response
sent by the service will be received with equal probability by all of its users,
and thus it is impossible to reliably provide auditors with a different set of
records without defeating the anonymizer.

We also posit the existence of some global channel that allows
a user to warn others that a fault has occurred, and that the adversary
cannot block.  We argue that this is a
legitimate assumption, since failures can be provided to third-party reporting
services or, if all else fails, manually sent via email to a public mailing list.  If
the server signs and time-stamps its responses, its misbehavior is non-repudiable,
thus preventing false-positives from being used to flood the channel.

\subsection{Probability of discord between pairs of users}
In our analysis, we consider two separate scenarios.  First, that where one user
wishes to verify the details of another without trusting that others clients will inform them
of inconsistencies in the responses.  This is the case with many legacy systems, for
example data from PGP keyservers or arbitrary websites, as it is reasonable to
assume that one might
be the only person attempting to verify the details of any particular user at any
given moment.

In the second scenario, the service acts like a traditional broadcaster---many users
attempt to access the same data, for example a Merkle tree root for
Certificate Transparency, CONIKS, or Bitcoin.  In this case we may assume that a certain
number of users are active in the protocol and able to publicly report failures---for
widely-distributed software, it is implausible that there would not be at least a few
hundred or thousand active users at any given time---allowing misbehaviour to be detected
with a yet-higher probability.

We start by considering the first case, where a given user is isolated from the
other users as in Figure~\ref{fig:results-diagram}.  Suppose $N$ users
each make $M$ identical requests to the sender-anonymous service.
It responds with $K$ copies of one message $x$, and $N-K$ copies
of another message $x'$.  These destinations of these
messages will be uniformly distributed over the set of users, as
shown in Theorem~\ref{thm:response-distribution}.

We begin by justifying our use of only two messages, $x$ and $x'$.

\begin{lemma}\label{lem:response-count}
When the described protocol is run with more than
two users, the maximum probability of successful equivocation occurs
when only two values are transmitted.

\begin{proof}
Suppose the service can transmit values $\{x, x', x'', \ldots\}$.  Then, for any choice of
responses, if $x'', x''', \ldots$ are replaced by $x'$, every sequence
of responses that do not trigger a failure by any set of users will still be accepted
by those users. Thus the maximum probability of successful equivocation is achieved
by a service transmitting only the true value $x$ and a single false value $x'$.
\end{proof}
\end{lemma}

This lemma is useful when bounding the probability of acceptance,
as it permits us to consider only two possible responses.  The goal
of the attacker, then, is that some users receive one response value
every time, and and others receive the
other response value each time.  Should Definition~\ref{def:anonymity}
hold, this is exceedingly unlikely, as we show in Lemma~\ref{lem:transfer}.

The analysis is eased substantially if we consider a perfect anonymization
system---that is, with $\epsilon = 0$---for which
the process of responding to the anonymous requests with one of two
responses is readily modelled by the process of pulling coloured balls from
an urn.  In this analogy, the response $x$ is represented by a white ball,
and $x'$ by a black one; the balls are drawn from the urn without replacement,
yielding the probability of a particular set of responses over the entire
set of users.

We begin by showing how a probability bound calculated with respect to a
$0$-sender-anonymous service can be loosened in order to apply to an
$\epsilon$-sender-anonymous service.

\begin{lemma}[{Imperfect anonymizer correction.}]\label{lem:correction}
Let $E$ be some event taken over the sample space
of $N!^M$ message$\rightarrow$recipient mappings for an $\epsilon$-sender-anonymous
service with $N$ users over $M$ rounds, and $\probsub{\epsilon}{E}$ be
the probability that $E$ occurs given some such service.

Then,
\begin{align}
   \probsub{\epsilon}{E} \le \probsub{0}{E} (1+\epsilon)^M , \label{eqn:correction}
\end{align}
where $\probsub{0}{E}$ is the probability that $E$ occurs
given a uniform distribution of mappings.
\begin{proof}
Let us first consider an individual outcome
\begin{align}
	r &\in \Omega = \mathrm{Bij}(\ZZ_N \rightarrow \mathcal{U})^M \\
		&= r_1 \times r_2 \times \cdots \times r_M ,
\end{align}
where the $r_i \in \mathrm{Bij}(\ZZ_N \rightarrow \mathcal{U})$ are the
response destinations for round $i$.  The event $\{r\}$ in which the outcome
$r$ occurs may then be written
\begin{align}
	\{r\} &= R_1 \cap \cdots \cap R_M
\end{align}
where
\begin{align}
	R_i = \; &\mathrm{Bij}(\ZZ_N \rightarrow \mathcal{U})^{i-1} \nonumber \\
				& \times r_i \nonumber \\
				&\times \mathrm{Bij}(\ZZ_N \rightarrow \mathcal{U})^{M-i-1} .
\end{align}
With a uniform probability measure $\probsub{0}{\cdot}$, $r$ occurs with probability $N!^{-M}$.

With our adversary-degraded probability measure $\probsub{\epsilon}{\cdot}$,
$r$ will occur with probability
\begin{align}
	\probsub{\epsilon}{\{r\}} &= \probsub{\epsilon}{R_1 \cap \cdots \cap R_M} \\
			&= \prod_{i=1}^{M} \condprobsub{\epsilon}{R_i}{R_1\cap \cdots \cap R_{i-1}} ,
\intertext{which Theorem~\ref{thm:response-distribution} bounds by}
			&\le \prod_{i=1}^{M} \frac{1}{N!} (1+\epsilon) \\
			&= (1+\epsilon)^M \probsub{0}{\{r\}} .
\end{align}
As the probability space is finite, we may write any event $E$ as a disjoint
finite union
\begin{align}
	E &= \bigcup_{e \in E} \{ e \} ,
\intertext{and thus}
	\probsub{\epsilon}{E} &= \probsub{\epsilon}{\,\bigcup_{e \in E} \{ e \}} \\
		&= \sum_{e \in E} \probsub{\epsilon}{\{ e \}} \\
		&\le (1+\epsilon)^M \sum_{e \in E} \probsub{0}{\{ e \}} \\
		&= (1+\epsilon)^M \probsub{0}{E} ,
\end{align}
the statement that we set out to prove.
\end{proof}
\end{lemma}

With this lemma in hand, we may bound the probability of any failure event
as though the anonymity of the users is perfect, applying a multiplicative
factor $(1+\epsilon)^M$ after the fact in order to account for the imperfect
nature of the anonymization system.  This is convenient for our calculations,
because the probabilities of our events of interest can then be easily determined
by straightforward coloured-balls-in-an-urn calculations.

\begin{lemma}\label{lem:transfer}
Suppose $N$
users take part in the described protocol exactly once, including a particular
pair of users Alice and Bob.
 The service provides $K$ copies of the message $x$, and $N-K$ copies
of a message $x'$.  Then, Bob will receive the value $x'$ 
and Alice the value $x$ with probability
\begin{align} p_i \le \frac{K (N-K)}{N(N-1)}  (1+\epsilon) . \end{align}
\begin{proof}
Where the service satisfies $\epsilon$-sender-anonymity with $\epsilon=0$,
the recipients of the responses are uniformly  distributed and thus
this problem is equivalent to that as that of pulling balls from an urn: given an urn
containing $K$ white
balls and $N-K$ black balls, acceptance is equivalent to drawing first a
white ball---probability \(K/N\)---and then a black ball---probability
\((N-K)/(N-1)\)---resulting in a joint probability of
\begin{align}p_{(b,w)} = \frac{K}{N} \frac{N-K}{N-1} .\end{align}
Application of Lemma~\ref{lem:correction} yields the original expression for
arbitrary security parameters $0 \le \epsilon$.
\end{proof}
\end{lemma}

\begin{theorem}[Probability of specific failure modes]\label{thm:transfer-m}

Suppose $N$
users take part in the described protocol with a sender-anonymous service,
including Alice and Bob.  Then, after $M$ iterations Alice will accept the value
$x$ and Bob the value $x'$ with probability at most
\begin{align}
	p_\mathrm{decep}
		\le \frac{\left[\frac{N}{2}\right]^M \left(N - \left[\frac{N}{2}\right]\right)^M}{N^M (N-1)^M} (1+\epsilon)^M.
\end{align}

\begin{proof}
We first consider the case where $\epsilon = 0$.
In order for Bob to accept a false value without detection by Alice,
the service must succeed all $M$ times in sending $x$ to Alice and some other
value $x'$ to Bob.  Lemma~\ref{lem:response-count}
indicates that the maximum probability of success occurs when only a single
false value is emitted, so we assume that all responses are either
$x$ or $x'$.

Suppose that in round $i$, the service responds $K_i$ times with $x$ and
$N-K_i$ times with $x'$. 

The probability that Alice receives $x$ and Bob $x'$ is given by
Lemma~\ref{lem:transfer} as
\begin{align} p_i \le \frac{K_i (N-K_i)}{N(N-1)} , \end{align}
and the probability of Bob accepting the false value without Alice noticing is therefore
\begin{align} p \le \frac{\prod_{i=1}^{M} K_i (N-K_i)}{N^M (N-1)^M} .\end{align}
This is maximized by setting $K_i = \left[N/2\right]$; when $N$ is odd, $K_i$ can be
rounded in either direction by the attacker---rounding up is more likely to result in the
true value being accepted, whereas rounding down increases the likelihood
of rejection.  The maximum probability of a successful attack is therefore
\begin{align}
	p_\mathrm{decep} \le \frac{\left[\frac{N}{2}\right]^M \left(N - \left[\frac{N}{2}\right]\right)^M}{N^M (N-1)^M} .
\end{align}
Application of Lemma~\ref{lem:correction} yields the original expression for
arbitrary security parameters $0 \le \epsilon$.
\end{proof}

\end{theorem}

As Alice and Bob are unaware of the number of other users accessing the service,
they must assume the worst-case value; this occurs when, $N=2$ yielding
$p_\mathrm{decep} \le 2^{-M}$.  As the number of users increases,
the bound will approach $4^{-M}$; we note again that this is the probability of
false acceptance for a single pair of users, and so does not take into
account the possibility that other users will detect the substitution and
report it publicly.

Theorem~\ref{thm:transfer-m} provides an important quantity that is directly
applicable to the security of a directory service: the maximum probability that
the service can deceive a user looking up a piece of information without being
noticed by a single auditor.  We can view this from another point of view,
namely the probability of breaking the consensus between pairs of nodes.
The essential difference is that Theorem~\ref{thm:transfer-m} does not
consider a broken consensus to be a failure if Alice accepts the value $x$, even
if Bob receives a copy of $x'$ and so reports misbehavior, despite the service having
successfully broken the consensus.

\begin{theorem}[Probability of pairwise discord]\label{thm:pair-consensus}
Suppose $N$ users take part in the described protocol, including Alice and Bob.  Then,
after $M$ rounds Alice and Bob will accept distinct values with probability at most
\begin{align}
	p_\text{\emph{split}} \le 2\frac{\left[\frac{N}{2}\right]^M \left(N - \left[\frac{N}{2}\right]\right)^M}{N^M (N-1)^M}  (1+\epsilon)^M.
\end{align}
\begin{proof}
We first consider the case where $\epsilon = 0$,
proceeding as follows: first, we calculate the probability that Alice and Bob will
receive different values in the initial round, then we apply Theorem~\ref{thm:transfer-m}
to calculate the maximum probability that they will both receive these initial values for
the remainder of the protocol.

Suppose that in the first round of the protocol, the server responds with
$K_1$ copies of the value $z$ and $N-K_1$ copies of the value $z'$.  Then,
the probability that Alice and Bob will receive different values is the probability
of receiving $z$ and $z'$ respectively, or $z'$ and $z$.  As these events are
disjoint, this probability is equal to
\begin{align}
	p_1 = 2\frac{K_1(N-K_1)}{N(N-1)}
\end{align}
by Lemma~\ref{lem:transfer}.

Let us denote the value received by Alice $x$ and that received by Bob $x'$.  Then,
by Theorem~\ref{thm:transfer-m},
the probability of that the remaining $M-1$ rounds will result in Alice receiving only
the value $x$ and Bob $x'$ is at most
\begin{align}
	\frac{\left[\frac{N}{2}\right]^{M-1} \left(N - \left[\frac{N}{2}\right]\right)^{M-1}}{N^{M-1} (N-1)^{M-1}} ,
\end{align}
yielding an overall consensus-breaking probability of
\begin{align}
p_\text{split}[K_1] \le \;2\,&\frac{K_1(N-K_1)}{N(N-1)} \nonumber\\
	& \times \frac{\left[\frac{N}{2}\right]^{M-1} \left(N - \left[\frac{N}{2}\right]\right)^{M-1}}{N^{M-1} (N-1)^{M-1}} .
\end{align}
This is maximized by setting $K_1 = [N/2]$ and thus
\begin{align}
p_\text{split} &\le 2 \frac{\left[\frac{N}{2}\right]^M \left(N - \left[\frac{N}{2}\right]\right)^M}{N^M (N-1)^M} .
\end{align}
Application of Lemma~\ref{lem:correction} yields the original expression for
arbitrary security parameters $0 \le \epsilon$.
\end{proof}
\end{theorem}

The probability of undetectably breaking the consensus between any pair
of nodes thus falls exponentially with time, never being greater than $2^{1-M}$.

\subsection{Probability of undetected consensus-breaking}\label{sec:broadcast}

We now take a more global view, and calculate the probability that the service can
equivocate without being detected by any of its users.  Where trustworthy reporting
infrastructure exists to allow the publication of equivocation reports to all of the users
of the service, this is the applicable probability of failure.  Furthermore, from the point
of view of the attacker or malicious service
operator, this is the probability that their attack will be detected, and thus the most
important consideration from a deterrance point of view.
The service may attempt to mislead as before, but the difficulty of achieving global
acceptance increased by the need to provide consistent responses to all users.

\begin{theorem}[Detection of consensus splits]\label{thm:consistency}
Suppose $N$ users take part in the described protocol, and the attacker
provides the response $x'$ to $K$ users and $x \ne x'$ to $N-K$ users.
If this is repeated over $M$ rounds, the probability that the attacker will
succeed in providing $MN$ consistent responses
to all $N$ users over all $M$ rounds is
\begin{align}
	p_c \le {N \choose K}^{1-M} (1+\epsilon)^M.
\end{align}
\begin{proof}
We first consider the case where $\epsilon = 0$.
	By Theorem~\ref{thm:response-distribution}, responses to an anonymous
	service are randomly assigned to users.  There exist $N\choose K$
	ways to assign the $K$ false responses amongst the $N$ users,
	and the attacker must do so identically to the first round for each
	of the $M-1$ subsequent or they will be detected.
	
	Note that $K$ must be the same for each round, otherwise at least
	one user will recognize the deception.

	This results in a non-detection probability of
	\begin{align}
	p_c \le \left[ {N \choose K}^{-1}\right]^{M-1} .
	\end{align}
	Application of Lemma~\ref{lem:correction} yields the original expression for
arbitrary security parameters $0 \le \epsilon$.
\end{proof}
\end{theorem}

This probability is maximized by setting $K$ as far from $N/2$ as
possible.  That is to say, a well-behaved (or consistently misbehaving)
service will respond consistently with probability $1$, and the maximum probability
of breaking the consensus between the users is $N^{1-M}$, achieved by providing identical
responses to all but a single user each round.

In addition to consistency checking, an attacker must contend with users
who have the ability to check the validity of their responses directly.  Should one
of these auditors receive the false value $x'$ directly, they can immediately
raise the alarm.

\begin{corollary}\label{cor:auditor}
If an auditor having knowledge of the true value $x$ takes part in the protocol,
then the probability of successfully
deceiving $K$ out of $N$
users---$N$ including the auditor---without detection by anyone is
\begin{align}
p_s[K] = \frac{N-K}{N}{N \choose K}^{1-M} (1+\epsilon)^M .
\end{align}
\begin{proof}
	We first consider the case where $\epsilon = 0$.
	We add an additional success criterion to Theorem~\ref{thm:consistency}.
	As well as responding consistently to each user, the service must respond to
	the auditor with the value $x$ in the first round.
	This occurs with probability $(N-K)/N$, and
	we multiply by the result stated in Theorem~\ref{thm:consistency}
	to obtain the result above.
	Application of Lemma~\ref{lem:correction} yields the original expression for
arbitrary security parameters $0 \le \epsilon$.
\end{proof}
\end{corollary}

\begin{theorem}\label{thm:transfer-group}
	The maximum probability of deceiving without detection \emph{any} member
	of a group of $N$ users, amongst them an auditor, who follow the
	protocol above for $M$ rounds is
	\begin{align}
		p_\mathrm{decep} \le \frac{N-1}{N^M} (1+\epsilon)^M .
	\end{align}
	\begin{proof}
		We first consider the case where $\epsilon = 0$.
		The value of $p_\mathrm{decep}$ above is that given by
		Corollary~\ref{cor:auditor} with $K = 1$.  We are only interested
		in the case where $K > 0$, since otherwise
		none of the group have been deceived, and with $K < N$,
		since then the auditor will detect the false message with
		probability one.  We write the bound from Corollary~\ref{cor:auditor}
		\begin{align}
			p_s[K] &= \frac{N-K}{N} {N \choose K}^{1-M} \\
			&= \frac{(N-K)(K! (N-K)!)^{M-1}}{N!^{M-1} N} ,
		\end{align}
		and hypothesizing that the maximum occurs
		when $K = 1$, we calculate
		\begin{align}
			\frac{p_s[K]}{p_s[1]}  &= \frac{(N-K)\left(K! (N-K)!\right) ^ {M-1}}{(N-1)(N-1)!^{M-1}} \\
			&= \frac{N-K}{N-1} \left(\frac{N}{{N \choose K}}\right)^{M-1} .
		\end{align}
		Since $0 < K < N$, both of these terms are at most one, and thus $p_s[K]$
		attains its maximum at $K = 1$, yielding the formula above, for $\epsilon = 0$.
		Application of Lemma~\ref{lem:correction} yields the original expression for
arbitrary security parameters $0 \le \epsilon$.
	\end{proof}
\end{theorem}

This demonstrates the difficulty of surreptitiously breaking the consensus between users shielded
by an anonymizer.  As before the probability of consensus-breaking falls rapidly
with protocol iterations, but this time the probability of deception approaches
zero---admittedly only polynomially---as the number of users increases.

\section{Anonymization methods}\label{sec:anonymization}

The question of how to perform the anonymization is not as straightforward
as it might first appear.  The simplest way is to use a mix-net, as this naturally
provides the lock-step behaviour that we have assumed in our analysis.
However, this infrastructure is not widely available, and so we briefly turn our
attention to more widely-deployed systems that might prove equally useful.

We use Tor in our prototype on account of its wide availability; in addition to its large
deployed capacity and mature software, its diversity of relay operators
renders systemic failure less likely than with a smaller-scale system intended
specifically for our protocol.

\begin{figure}
	\centering
	\includegraphics{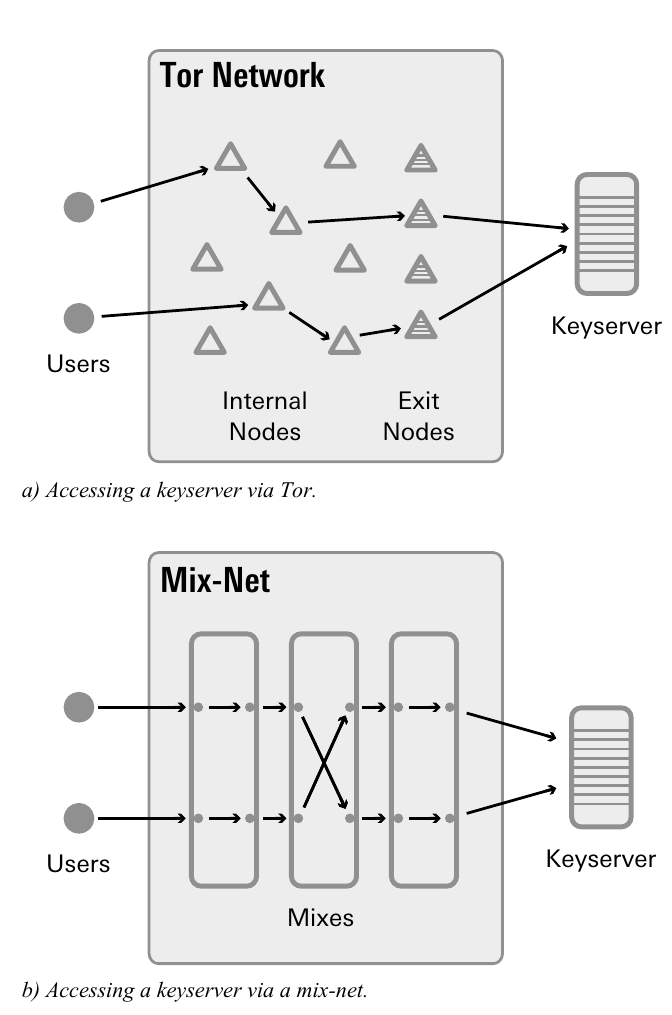}
	\caption{Connecting to a public keyserver via Tor and via a mix-net.
			The user randomly selects
			several relays, then adds a layer of encryption for each relay.
			After receiving a message, the relays strip their layer of encryption,
			revealing the address of the next relay.  Eventually, the message
			reaches an \emph{exit node}, which passes it to the open
			internet.
			Anyone can contribute nodes to the network---including
			adversaries---however as the routing path is selected by
			the user, an attacker cannot gain access to the encrypted
			messages with probability better than chance.
			Mix-nets are composed of a chain of \emph{mixes}, which
			take batches of messages, remove a layer of encryption,
			shuffle the messages, then pass them to a new mix.  If at least
			one mix in the chain is honest, then an attacker cannot connect
			messages to their senders with probability better than chance.}
	\label{fig:tor-public}
\end{figure}

Rather than transmitting batches of messages in lock-step, as a mix-net
does, Tor immediately forwards its received messages---termed
\emph{onions} for their layers of encryption---to the following relay
or, if they are the last in the routing chain, to their destination.  This
reduces the latency of the system, making it usable for interactive tasks.
The difference between the structures of these two systems is shown in
Figure~\ref{fig:tor-public}.

Despite this, while Tor may render difficult the task of determining
which sites a user visits, or conversely which users are visiting a site,
our requirement of anonymity at the level of individual requests is more
difficult.  The first and most obvious point is that Tor channels are reused
for ten minutes at a time, and therefore client software must demand
a new channel for every request in order to prevent them from being linked
by IP address.  Even so, users must be exceedingly careful if
they are to avoid giving information away via fingerprinting of their
client software.

Another risk is that information will be leaked via timing attacks; if the requests
are made at a fixed time, then the order in which the server receives the
requests may allow it to link the requests by the clock error of each
user.  The time of each request must therefore be randomized,
as must the times at which channels are set up.  An important topic for
future work is therefore to develop an asynchronous alternative
to the protocol that we have described.

Client software poses a risk as well---if the service being audited
uses HTTPS, it might attempt to fingerprint a user by its
available cipher suites, or by the time needed for negotiation to
take place.

Tor's use of a long-term guard relay substantially degrades short-term linkability,
despite its utility in maintaining anonymity over the long term.  Guard relays are
stable relays that are selected by the client and then used as the first hop over
a period of weeks to months before being changed~\cite{elahi-changing-of-the-guards}.
If the clients do not use a long-term guard relay, then they become vulnerable to
predecessor attacks~\cite{predecessor-attack}, in which a malicious relay simply
waits until it is selected as the first hop by the client, which it can recognize
with traffic analysis.  

Our concern is that an attacker will be chosen as a
guard with non-negligible probability, effectively guaranteeing that that client will be deceived,
and reducing $N$  by one in the previous analysis.  To avoid this,
the first hop must change with every request, requiring reconfiguration of Tor.

A possibly more secure approach would be to use some kind of protocol that responds to a
single fixed datagram packet, however as Tor does not support UDP,
this approach would require the use of some other anonymizing network.
Nonetheless, with careful design it will be possible to reduce the
information leakage to a level that sufficiently masks the source of each
request.

Another approach is to use auditing servers, as suggested in the CONIKS
architecture~\cite{coniks}.  This is similar to Tor in many respects, with
clients selecting the server from which their traffic will appear to come.  The use
of dedicated auditing servers has some advantages in that they can
sign responses from the server, allowing a degree of undeniability on
the part of the service being audited, at least to the extent that
the auditing servers are trusted.  In addition, the auditing server can cache responses from the
service, reducing its load and forcing it to commit to its equivocated response
for all subsequent requests made to that auditor.  This comes at the cost
of new server infrastructure with multiple independent operators, or
equivalent changes to existing anonymizing systems; this prevents the technique
from being immediately useful, however the security gain achievable by such
`intelligent' systems is a worthy avenue for future work.

\subsection{Assignment of trust}

Given that this protocol requires a trustworthy anonymity system, one might
reasonably ask what has been gained from a trust point of view.  We avoid the
need for a Byzantine consensus protocol, but depend upon systems like
Tor whose basic security properties are themselves dependent upon a consensus
protocol.

While such infrastructure might be used directly to audit the service in question,
as discussed in previous works, this creates a bootstrapping problem.  The value
of a consensus protocol derives from the fact that users believe that most
of participants are honest.  The operators of the Tor directory operators are
trusted by the community, the directories that they produce are small enough
to be well-scrutinized, and the consequences of misbehaviour are large.
The result is that Tor is---to most users---more trustworthy than any new
auditing mechanism will be.

The protocol that we describe in this paper
does not allow us to completely sidestep the need to trust
an infrastructure provider, but rather allows trust to be restricted to
third parties that have no particular interest in the system in question.
This solves the current problem of unavailability of trustworthy participants
to emerging systems.

\subsection{Failure of the anonymity system}

Onion routing sacrifices some of the security of mix-networks
for low latency~\cite{tor-design}.  Despite the vulnerability to traffic analysis that results,
low latency allows the system to be used for web browsing and other real-time applications,
and has driven Tor's wide adoption.  Our anonymity system model results in
a loose security reduction, the value $\epsilon$ increasing rapidly with
the number of users.

We consider two quite similar cases: a mix-net with a single honest mix whose outputs are
surveilled, and the Tor network.  In both of these cases, each request is deanonymized
with a fixed probability and independently of the other requests.

We calculate $\epsilon$ as follows: let $D$ be the number of users that have been
deanonymized; this is binomially distributed, with distribution $\mathrm{Bin}(N, p_d)$.
The attacker has no knowledge of which requests belong to the other users, and must
therefore guess them; this is successful with probability $(N-D)!^{-1}$.  Combining these,
we compute the probability of correctly identifying the source of every request:
\begin{align}
	&\prob{\expsa_{\mathcal{L}_\mathrm{Tor},\mathcal{U},1}(\mathcal{A}) = 1} \\ &= \frac{1}{N!}\left(1+\epsilon\right) \\
		&= \sum_{d = 0}^{N} \prob{D=d} \frac{1}{(N-k)!} \\
		&= \sum_{k = 0}^{N} {N \choose k} p_d^k (1-p_d)^{N-k} \frac{1}{(N-k)!}
\end{align}
and thus
\begin{align}
	 \epsilon
		 &= -1 + \sum_{k = 0}^{N} p_d^k (1-p_d)^{N-k} \frac{N!^2}{(N-k)!^2 k!} \label{eqn:epsilon-torlike}.
\end{align}

In the case of a mix-net, deanonymization occurs when a layer of encryption is broken
for a message; this allows the content at the input and output of the honest mix to be
linked.  Arbitrarily setting $p_d = 2^{-80}$ and $N=10$, we obtain
$\epsilon \approx 2^{-79}$; this is essentially negligible. Even with $10^4$ users,
$\epsilon$ increases to a still-negligible $2^{-53}$.

With Tor,
we examine the least favourable case, that where the service in question is malicious.  This means
that the attacker has knowledge of the anonymized request times and control over the
responses.  Traffic analysis allows them to deanonymize a request whenever one of their
relays is selected by the client for the first hop, thus revealing the client IP address.  Suppose that
there are $N_i$ entry relays, of which $C_i$ are compromised or surveilled by the attacker.  Then,
if relays are chosen uniformly, the attacker can deanonymize a channel with
probability $p_d = C_i/N_i$.  In reality, modern versions of Tor do not select relays
with uniform probability, but weighted by bandwidth~\cite{elahi-changing-of-the-guards};
this can be accounted for by defining $C_i$ and $N_i$ to be bandwidths rather
than node counts, however~\cite{sybilhunter} does not provide this information
for the cabals that they detected.

If we consider a reasonably large cabal of $100$ malicious relays out of $7000$, for $2$ users,
Eqn.~\ref{eqn:epsilon-torlike} yields the small but non-negligible $\epsilon = 0.003$.  This quickly increases to
$\epsilon=2.13$ for only $10$ users, a substantial loosening of the security
bound.  

While we are constrained by space from re-deriving
all the results above with respect to the properties of Tor,
we will derive an equivalent to Theorem~\ref{thm:transfer-m} for
a low-latency onion router, specifically Tor.  This demonstrates that control of a moderately-sized
cabal of Tor relays does not greatly reduce the security of our initial prototype
relative to what we have proven above.

\begin{theorem}\label{thm:transfer-m-onion}
Suppose $N$ users, including Alice and Bob, access a service via an onion router,
each of them being deanonymized independently with probability $p_d$ during each round.
Each user executes the protocol described in Section~\ref{sec:protocol} for $M$
rounds.  Then, the probability that Alice will consistently
receive the response $x$ and Bob $x'$ is bounded as
\begin{align}
	p_\mathrm{decep} \le \left[ 1 - \frac{1}{2}(1-p_d)^2 \right]^M.
\end{align}

\begin{proof}
For each round, three cases are possible: neither Alice nor Bob are
deanonymized, occurring with probability $(1-p_d)^2$, or one is
deanonymized, this time with probability $2p_d(1-p_d)$, or both are
deanonymized, this occurring with probability $p_d^2$. In the last case, the attacker's
success is trivial for that round.  The same is true if only one of the pair are
deanonymized---we suppose
without loss of generality that it is Alice---because the server can respond to Alice with
$x$, and to everyone else with $x'$.

If neither Alice nor Bob have been deanonymized, Theorem~\ref{lem:transfer}
applies, with the number of users $N_a \ge 2$ being that remaining in the anonymity set.  The
probability of deception is therefore
\begin{align}
p_r &= \frac{K(N_a-K)}{N_a(N_a - 1)} \\ &\le \frac{\frac{N_a}{2}\left(N_a-\frac{N_a}{2}\right)}{N_a(N_a - 1)} \\
	&= \frac{N_a^2}{4 N_a(N_a-1)}\\
	&\le \frac{1}{2} .
\end{align}

This occurs with probability $1-(1-p_d)^2$, and thus the maximum probability that
the attacker succeeds during a given round is 
\begin{align}
	&1-(1-p_d)^2 + (1-p_d)^2  p_r \\ =\; &1 -(1-p_d)^2 (1-p_r) \nonumber \\
		\le \; &1 - \frac{1}{2}(1-p_d)^2.
\end{align}
Success in each round is independent, so this occurs $M$ times with
probability
\begin{align}
	p_\mathrm{decep} \le \left[ 1 - \frac{1}{2}(1-p_d)^2 \right]^M.
\end{align}

\end{proof}
\end{theorem}

Using Theorem~\ref{thm:transfer-m-onion}, this yields a deception probability
\begin{align}
	p_\mathrm{decep} \le \left( 1-\frac{1}{2}\left(1-\frac{C_i}{N_i}\right)^2 \right)^M .
\end{align}

From \cite[Table~2]{sybilhunter}, we see that most malicious relay groups
which escape detection for any length of time have less than $100$ members.
The Tor network, by comparison, has approximately $7000$ relays~\cite{tormetrics}
at the time of writing.  The effect is to loosen the bound on attacker success from
$p_\mathrm{decep} \le 0.5^{M}$ to $p_\mathrm{decep} \le 0.514^M$.  This shows that
Tor achieves the original $N=2$ security bound with an arbitrarily large number of users, and
so remains useful despite its poor $\epsilon$-values in our more general analysis.

\section{Discussion}

We have presented a protocol that uses an anonymizing service to create
an auditable broadcast service.  This capability is extremely valuable, and can
be used in several ways.  We have demonstrated how anonymizing networks
can be used by individuals in order to distribute their own public keys, but
with more and more systems being designed to allow the verification of database
entries via a Merkle tree~\cite{laurie-transparency,keybase,coniks}, we
must analyze this type of system as well.
In this case, many users can be assumed to
access the same service simultaneously, and therefore the results from
Section~\ref{sec:broadcast} apply.  If more than a handful of users take part
then detection is near-certain, even with very few rounds.

The requirement that the holder of an identity takes part is an onerous one,
but one that could be met should such a technique become ubiquitous, for
example if it is performed automatically by default installations of PGP
implementations by all major vendors.  Even if this were not the case, the
approach still serves to reassure the holder of an identity that other users
can communicate securely with them if they choose to take this approach.

The need for multiple rounds makes this approach relatively expensive in
terms of communication.  This, in addition to the time needed for
failure reporting, rules it out in most interactive applications.  With systems like
CONIKS this is not a problem, as data that is a few minutes out of date
will not cause any great harm, since the data being broadcast allows
any user to be looked up.  When verifying individual keys using the
existing PGP keyserver network, the process must be performed separately
for each key.  This results in an delay before first communication can take
place, but subsequent verification can be performed in the background to
ensure that the previously-verified key is up to date.

\subsection{Implementation analysis}
Our discussion thus far has been quite general, and we briefly discuss what
can be achieved in practice.

The relevant parameters for the system when used to audit keyservers and
Merkle Trees is shown in Table~\ref{tbl:results}.
\begin{table}[t]
	\renewcommand{\arraystretch}{1.4}
	\centering
	\begin{tabular}{c | c c}
		\hline
		&\shortstack{Single\\Records} & \shortstack{Merkle Tree \\Root} \\
		\hline
		Number of users assumed & $2$ & $N \gg 2$\\
		\hline
		Items validated per user & $L$ & $L$ \\
		\hline
		Number of requests per user & $ML$ & $M+L$ \\
		\hline
		Probability of undetected failure & $2^{-M}$ & $(N-1)N^{-M}$ \\
		\hline
		Legacy system support & Yes & No \\
		\hline
	\end{tabular}
	\caption{Costs and security of the proposed protocol for
			literal-data and Merkle Tree systems.}\label{tbl:results}
\end{table}

The first scenario that we consider is the verification of entries on a
PGP keyserver.  In this case users access keys directly, verifying them
on an individual basis.  It is necessary to make a trade-off here between
the time needed to achieve a reasonable level of verification, and
the load placed upon the keyserver.

The requests in this case take the form of search queries for the
email address in question.  Both the users and the identity holder must
agree on the form of these search queries and how the the key is to be
selected from the results.  In our implementation, the search query is
an email address, and the result is taken to be the first valid key listed in
the response.

We suppose here that a round will take place every $T = 5$~minutes; thus
after time $t$, $M = \lfloor t/T\rfloor$ rounds will have taken place.
Therefore, in order to achieve a maximum failure probability $p_\text{decep}$, we
require a verification time
\begin{align*}
	t = - T \log_2 p_\text{decep} .
\end{align*}
If we arbitrarily determine a success probability of $2^{-20}$ to be
reasonable---it seems implausible that we could do substantially better
by any other means, including in-person verification of identity
documents---then verification requires $100$~minutes,
with server load being inversely proportional to the verification
time.  This is somewhat inconvenient, but far less so than
obtaining a personal certification, which in all likelyhood will
require several hours of time in order to coordinate, travel,
and perform the verification.

Next we consider the Certificate Transparency system.  This requires
that a user periodically obtain a Merkle-tree root, with newer roots
attesting to previous values as well.  We model our system on
Chrome's software-update system, supposing that the root will be
downloaded at the same time.  Chrome checks for software
updates every five hours~\cite{google-chrome-updates}; if it were
to randomize the time of checking during each five-hour interval,
then this matches the situation that we have analyzed, with the
obvious exception being that Chrome does not currently perform any
anonymization.

\begin{figure}[h]
	\centering
	\includegraphics[width=58mm,height=57.53mm]{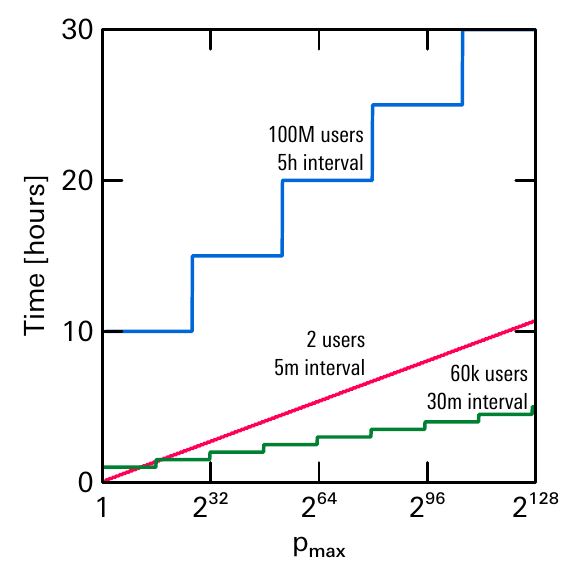}
	\caption{Waiting-time necessary to achieve various
			levels of security.  We show the hypothesized
			Certificate Transparency system modelled on
			the Chrome auto-update mechanism (top),
			our proposed keyserver-auditing system (middle),
			and our conception of how a keyserver
			built on something like CONIKS might look (bottom).
			We see that very small probabilities of equivocation
			are achieved within only a few hours, such that
			deanonymization and endpoint compromise quickly
			become far more likely than chance success by
			a malicious service.}
	\label{fig:delays}
\end{figure}

We make a conservative estimate of Chrome having
$100$~million active users, though in reality it is most likely
several times higher.  This time we have $T = 5$~hours,
and from Table~\ref{tbl:results} we will require
\begin{align*}
	t = - T \frac{\log \frac{p_\text{decep}}{N-1}}{\log N} .
\end{align*}
In this case, it is straightforward to obtain a probability of
deception of at most $2^{-20}$---after the second request,
the probability that anyone will be deceived without
the misbehaviour being detected by at least one browser
instance or site owner is $2^{-26}$, or approximately
$1$ in $100$~million.  These waiting times are shown in
Figure~\ref{fig:delays}.

We reinforce here that this probability is the
maximum probability that the service may succeed
in deceiving \emph{any} user.  Thus the average number
of users deceived is approximately $p_\text{decep}$---it is
possible, albeit unlikely, that more than one user will be deceived---and
\emph{not}
$N p_\text{decep}$.

We see, then, that our results are useful in practice and can
provide meaningful security against malicious services.

\subsection{Effects on the Tor network}

Our somewhat unusual use of Tor raises the important question of whether the
use of Tor in our system poses a risk to other users of the network, or
conversely whether it might improve the anonymity provided by Tor.
Our need to disable entry guards disabled renders clients using our protocol
is rather distinctive, but it is not clear whether this is problematic.

A greater risk from a usability perspective is that misconfigured applications might use
our unusually-configured version of Tor for traditional applications, leaving
users vulnerable to predecessor attacks.  This might be avoided through
application-filtering by a local firewall, but safest of all is to use a modified Tor
client that enforces some kind of client authentication.

A potential positive effect of this protocol is the enlargement of the
anonymity set of Tor users, though this must be balanced against the
ease with which an eavesdropper can differentiate between Tor users
using our protocol and those using Tor in a more traditional manner.  Because
the protocol is not highly latency-sensitive, a hypothetical onion router
that allows clients to request some delay before
the packet is retransmitted might reduce the risk of traffic confirmation attacks
to the point that the use of an entry guard can be used, thereby making
the use of our protocol far less obvious.
\section{Conclusion}

We have shown how an anonymizing service such as Tor can be used to perform
multi-path probing, and so create
a public broadcast channel that permits clients to bound the probability that the
broadcasting service can break consensus with the other clients without detection.
Failed attempts to provide different messages to
different parties can be proven by the detecting party with the aid of digital
signatures.  Such a protocol has the potential to provide remote verification of
public keys, rendering end-to-end public-key cryptography possible without the
need for trust in certificate authorities or for potentially insecure approaches such
as trust-on-first-use.  We have bounded the probability that an equivocating
service will succeed in deceiving its users, and have provided a security reduction
to the anonymization-resistance of the underlying anonymization service.

\bibliographystyle{IEEEtran}
\bibliography{E:/Documents/docear-saved.bib,E:/Documents/rfc.bib}

\appendix
\section{Implementation}\label{sec:impl}
We have developed an implementation of this system, which we have dubbed
\emph{Keywatch}\footnote{\url{https://github.com/LachlanGunn/Keywatch}}.  It takes the form of a curses
program that continuously searches for a number of email addresses on an
HKP keyserver~\cite{hkp} via Tor.  We chose to use Tor rather than
a mix-network because of its wide public availability.

Requests are made via \emph{libcurl}, using Tor's authentication isolation
feature~\cite[\emph{IsolateSocksAuth}]{tor-torrc-documentation} feature to force the creation of new
channels.  Connections are made using plain HTTP, reducing the potential
for fingerprinting by the client's TLS configuration and round-trip times.
The client is relied on to have a sufficiently accurate clock, which is used
to determine the time window for each round of the protocol.  The windows
are \SI{10}{\second} in duration, and defined to start at integer numbers
of periods since 2000-01-01 0000 GMT. This duration is short and only suitable
for testing; before leaving the prototype stage, it will be lengthened to
several minutes.

Since the clocks of the clients are not necessarily well-synchronized, 
the request times allow fingerprinting of the clients.  In order to avoid this,
the time of each request within each window is chosen at random according
to $X_n T /{2^{64}}$, where $T$ is the window duration and $X_n$ is a
random number between $0$ and $2^{64} - 1$ found
by filling a 64-bit unsigned integer with bytes from the operating-system
cryptographic random-number generator.

After the index is downloaded, the fingerprint associated with the email
address is taken to be that of the first valid---that is to say, unrevoked
and unexpired---key to which the email address is associated.  Once such
a fingerprint has been received, it is retained in memory and compared
with the first valid fingerprint from each subsequent request.  Should they
differ, the key provided by the offending request will be printed to the
terminal.

\end{document}